\documentclass[aps,prb,reprint,amssymb, amsmath, superscriptaddress, longbibliography]{revtex4-2}

\usepackage{lipsum, xcolor,graphicx, amsmath}
\usepackage{placeins, float}
\usepackage{amssymb}
\usepackage{braket}
\usepackage{bm}
\usepackage[T1]{fontenc}
\usepackage[colorlinks=true, citecolor={blue!80!black}, urlcolor={blue!50!black}, linkcolor = {blue!80!black}]{hyperref}

\begin{document}

\title{Low-field magnetization processes of hexagonal easy-plane altermagnet $\alpha$-\textrm{MnTe}}


\author{Sahana R\"{o}{\ss}ler}
\email[]{sahana.roessler@uni-leipzig.de}
\affiliation{Felix Bloch Institute for Solid-State Physics, University of Leipzig, D-04103 Leipzig, Germany}
\author{Victoria Ginga}
\affiliation{Felix Bloch Institute for Solid-State Physics, University of Leipzig, D-04103 Leipzig, Germany}
\author{Marcus Schmidt}
\affiliation{Max Planck Institute for Chemical Physics of Solids, N{\"o}thnitzer Str. 40, 01187 Dresden, Germany}
\author{Yurii Prots}
\affiliation{Max Planck Institute for Chemical Physics of Solids, N{\"o}thnitzer Str. 40, 01187 Dresden, Germany}
\author{Helge~Rosner}
\affiliation{Max Planck Institute for Chemical Physics of Solids, N{\"o}thnitzer Str. 40, 01187 Dresden, Germany}
\author{Ulrich Burkhardt}
\affiliation{Max Planck Institute for Chemical Physics of Solids, N{\"o}thnitzer Str. 40, 01187 Dresden, Germany}

\author{Ulrich K. R\"{o}{\ss}ler}
\affiliation{Institute for Theoretical Solid State Physics, Leibniz IFW Dresden, 01069 Dresden, Germany}

\author{Alexander A. Tsirlin}
\affiliation{Felix Bloch Institute for Solid-State Physics, University of Leipzig, D-04103 Leipzig, Germany}

\date{\today}

\begin{abstract}

%

Single crystals of $\alpha$-MnTe were synthesized by chemical vapor transport using iodine as the transport reagent. Structural characterization by powder x-ray diffraction confirmed the hexagonal structure (space group P6$_{3}$/mmc). Magnetization $M(T)$ and specific heat $C_p(T)$ measurements revealed an antiferromagnetic phase transition at $T_N \approx307$ K. The magnetic entropy derived from the $C_p(T)$ data is consistent with the $S = 5/2$ spin state of Mn$^{2+}$ ions. Angle- and field-dependent magnetization measurements indicate complex magnetic responses associated with domains, and show an anomaly around 1 T. These features are analyzed using a phenomenological micromagnetic model that includes higher-order anisotropic exchange interactions coupling the weak ferromagnetic component and the antiferromagnetic order parameter. The model captures the generic behavior of magnetic states and demonstrates that the observed uniaxial and unidirectional anisotropies arise from metastable domain configurations and irreversible magnetization processes.    

\end{abstract}


\maketitle

\section{Introduction}

Antiferromagnetic compounds that display non-relativistic spin-polarized band splitting due to crystal symmetry have been newly classified as altermagnets \cite{Lib2022a,Lib2022b}. They have gained traction owing to their interesting fundamental properties and their potential applications in antiferromagnetic spintronics. In this context, several compounds such as, RuO$_{2}$, $\alpha$-MnTe, and CrSb, have been the subject of intense investigation \cite{Bai2024}. However, it has since been shown that RuO$_{2}$ is non-magnetic at least in bulk form \cite{Hir2024,Kesler2024,Kie2025,Liu2024,Max2025}. In contrast, $\alpha$-MnTe exhibits several characteristics of an altermagnet, making it a prime example of this class of compound. It is a doped semiconductor with $g$-wave symmetry and has the N\'eel temperature $T_N \approx 307$\,K, which may be sufficiently high for room-temperature applications. The lifting of Kramers degeneracy, one of the defining features of altermagnets, has been observed in angle-resolved photoemission spectroscopy (ARPES) experiments on both thin films \cite{Lee2024,Krem2024,Haj2024} and bulk single crystals \cite{Krem2024,Osumi2024}. These studies report a spin-splitting magnitude of 0.2–0.4 eV at low-symmetry momentum points. A combined study by inelastic neutron scattering and model based spin-wave theory calculations observed a split magnon bands of the magnitude 2\,meV \cite{Liu2024a}. Further, sophisticated spectroscopic imaging technique have been utilized to map the altermagnetic domains in thin films \cite{Amin2024} and single crystalline lamella \cite{Yama2025} of $\alpha$-MnTe. 

Although altermagnetic properties of $\alpha$-MnTe are now well established, several fundamental questions remain unanswered. For example, it is expected that altermagnets will exhibit a spontaneous anomalous Hall effect (AHE) without a net magnetization, via the Berry curvature mechanism \cite{Libor2022}. Spontaneous AHE in the absence of weak ferromagnetism has been reported in $\alpha$-MnTe thin films \cite{Beta2023,Liu2025}. But it has been also shown that AHE depends strongly on the substrate chosen for growing $\alpha$-MnTe thin films as well as the cooling history \cite{Bey2024} thereby deviating from the behavior expected from the theory. Furthermore, in bulk crystals, AHE was observed in the presence of weak ferromagnetism, weak ferromagnetism that appears below $T_N$ \cite{Kluc2024}. Further, spontaneous magnetization appearing at 81 K in $M(T)$ measurements have been reported \cite{Orlova2025}. 

The origin of the weak ferromagnetism has been attributed to chiral biquadratic
interaction, i.e., a type of higher order Dzyaloshinskii-Moriya (DM) interaction \cite{Mazin2024}. Alternatively, it has also been attributed due to non-collinear $g$-tensor \cite{Jo2025}. Both effects are induced by spin orbit coupling. However, it is clear that parasitic weak ferromagnetic moments can also be caused by vacancies and defects \cite{Chil2024}. In the latter case, the weak ferromagnetism has been linked to the Mn-richness, which is inherent to $\alpha$-MnTe thin films grown by MBE \cite{Chil2024}. Indeed, $\alpha$-MnTe is not a line compound, but rather has a homogeneity range for $42.5-51\%$ Te, which is reported in the Mn--Te binary phase diagram \cite{Schel1998}. Deviations from the 1:1 stoichiometric ratio can also affect the lattice parameter, electrical resistivity, and carrier concentration. A careful control of the stoichiometry is therefore required to understand the magnetic behavior of $\alpha$-MnTe and to elucidate the effect of the chemical composition. Single crystals used in previous studies were typically grown from the melt and might contain compositional gradients and higher defect densities. To this end, we report single crystals of $\alpha$-MnTe grown by chemical vapor transport. This method allows the growth of crystals with fewer defects, higher purity, and better stoichiometry than melt-grown crystals. We present resistivity, specific heat, as well as field-, temperature-, and angle-dependent magnetization. The peculiar uniaxial and unidirectional anisotropic behavior observed in magnetic fields is analyzed using a phenomenological model considering higher-order anisotropic exchange with linear-cubic anisotropic invariants between weak ferromagnetic magnetization and the N\'eel vector.



\section{Experimental methods}
The compound $\alpha$-MnTe was synthesized and crystallized in a two-stage process, based on the procedure described in \cite{Melo1991}. First, polycrystalline $\alpha$-MnTe was synthesized by a direct reaction of the elements in an equimolar ratio of Mn (pieces, Chempur 99.99 \%, powdered directly before use) and Te (powder, Alfa Aesar 99.999\%) with the addition of iodine (Alfa Aesar 99.998 \%) at 500 °C in evacuated and sealed quartz glass tubes over a period of 10 days.  Subsequently, starting from the synthesized polycrystalline sample, crystals of $\alpha$-MnTe were grown by chemical transport in a temperature gradient from 700 °C (source) to 650 °C (sink) with the addition of 1.5 mg/ml iodine (Alfa Aesar 99.998 \%) as a transport agent. The crystals were typically grown over a perieod of 10 days. The ampules containing single crystals were cooled down by quenching in water kept at room temperature. The selected crystals were characterized by X-ray diffraction on powder and single crystal as well as by EDXS and back-scatter electron (BSE) imaging.

\begin{figure}[t]
    \centering
    \includegraphics[clip,width=0.95\columnwidth]{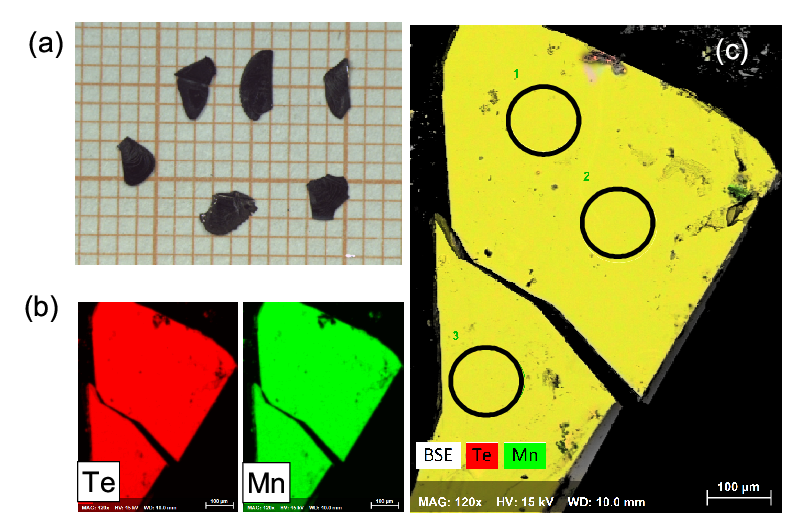}
    \caption{(a) An image of single crystals grown by chemical vapor transport. (b) Back-scatter electron (BSE) map of a $\alpha$-MnTe single crystal, individually color-coded for Mn and Te ions. (c) BSE map of $\alpha$-MnTe. The circles represent the area in which the composition was analyzed using EDXS. At positions marked 1,2, and 3, the compositions were consistently found to be Mn$_{1.02(1)}$Te$_{0.98(1)}$, Mn$_{1.02(1)}$Te$_{0.98(1)}$, and Mn$_{1.02(1)}$Te$_{0.98(1)}$, respectively.} 
    \label{fig:EDX}
\end{figure}
 
For high-resolution synchrotron x-ray powder diffraction measurements, fine powder of $\alpha$-MnTe was filled in a quartz capillary that was spun during the experiment. The powder sample was prepared by grinding single crystals. The measurements were performed at the ID22 beamline of the ESRF (Grenoble, France) with a wavelength of 0.35433 {\AA} at 298 K using N$_{2}$-flow cryostream. Structural model was refined by the Rietveld method using the program JANA2006 \cite{Petr2014}. The background parameters were fitted using Legendre polynomial function, and the peak shapes were described by a Pseudo-Voigt function. 

Temperature and field dependence of magnetization was measured in a SQUID vibrating sample magnetometer (MPMS3, Quantum Design) in the temperature range 370 -- 2\,K and in magnetic fields up to 7\,T. The angle dependence of magnetization $M(\phi)$ was conducted in a SQUID magnetometer (MPMS XL, Quantum Design) using a mechanical rotator. The heat capacity $C_p(T)$ and resistivity $\rho(T)$ measurements were carried out in the Physical Property Measurements System  (Quantum Design).   

\section{Results and discussion}
\subsection{Experimental characterization}
\begin{figure}[t]
    \centering
    \includegraphics[clip,width=0.95\columnwidth]{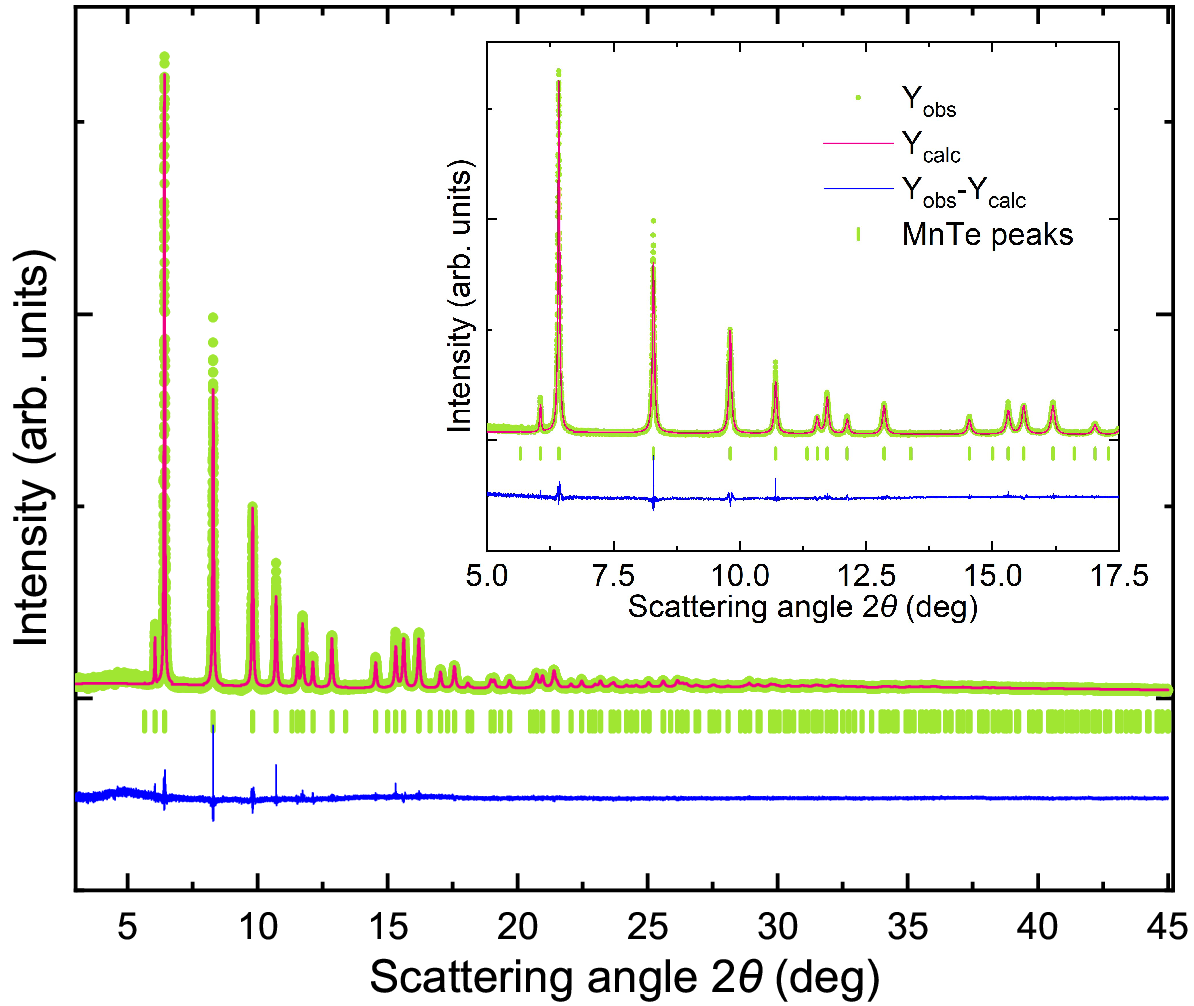}
    \caption{Results of the profile matching by Rietveld refinement of $\alpha$-MnTe from the powder x-ray diffraction data taken at 298 K.}
    \label{fig:XRD}
\end{figure}

  Fig.\,\ref{fig:EDX}(a) shows plate-like single crystals of $\alpha$-MnTe obtained by chemical vapor transport. The images of the BSE maps presented in Fig.\,\ref{fig:EDX}(b,c) display a homogeneous distribution of Mn and Te throughout the crystal.  No clustering or secondary phases were observed in the BSE maps. EDXS analysis revealed that the crystal's composition deviated slightly from a 1:1 stoichiometric ratio, with an average excess up to 2\,\% of Mn. The excess Mn ions are expected to occupy the tetrahedral holes of hexagonal close packing of Te ions. This type of non-stoichiometry is typically found in NiAs-type structures \cite{Kjekshus1964}. The powder x-ray diffraction of the crystals confirmed NiAs-type structure with space group $P6_3/mmc$ (No. 194). Further, high-resolution XRD was used to confirm the absence of symmetry lowering. 
The refined unit-cell parameters for $\alpha$-MnTe are: $a\,=\,$4.1450(5) {\AA} and $c\,=\,$6.7105(2) {\AA}. The final observed and calculated powder XRD patterns are given in Fig.\,\ref{fig:XRD}. The parameters of the full-profile Rietveld refinement for $\alpha$-MnTe are presented in Table \ref{tab1}. The atomic coordinates and isotropic displacement parameters are given in Table \ref{tab2}.   

\begin{table}[t]
\caption{The parameters of the full-profile Rietveld refinement of powder x-ray diffraction data for the powdered single crystal sample of MnTe.} 
\label{tab1}
\begin{ruledtabular}
\begin{tabular}{ll}
Temperature  & 298~K \\
\hline
Space group      & $P6_3/mmc$  No. 194  \\
 \hspace*{0.28cm}$a$ (\AA) & 4.1450(5)              \\
 \hspace*{0.28cm}$c$ (\AA) &  6.7105(2)    \\
 \hspace*{0.28cm}$\lambda$ (\AA)	& 0.35433  \\
 \hspace*{0.28cm}Cycles of refinement	& 30 \\
  \hspace*{0.28cm}Step ($^{\circ}$) & 0.010  \\
	\hspace*{0.28cm}Profile function & Pseudo-Voigt \\
	\hspace*{0.28cm}$R_p$ (\%)& 2.3 \\
	\hspace*{0.28cm}$R_{wp}$ (\%)& 3.41  \\
	\hspace*{0.28cm}$R_{exp}$ (\%)  &2.36 \\
	\hspace*{0.28cm}GoF & 1.06 \\	
\end{tabular}
\end{ruledtabular}
\end{table}
\begin{table}[t]
\caption{Atomic coordinates, and isotropic displacement parameters for $\alpha$-MnTe.} 
\label{tab2}
\begin{ruledtabular}
\begin{tabular}{lllll}
Atoms  &x &   y    & z    & $U_{eq}$ (${\AA}^2$)\\
\hline 
Mn &0 &   0    & 0   & 0.0165(2) \\ 
Te &  1/3  &2/3   & 1/4  & 0.00871(9)\\
\end{tabular}
\end{ruledtabular}
\end{table}
%
%
\begin{figure}[t]
    \centering
    \includegraphics[clip,width=0.95\columnwidth]{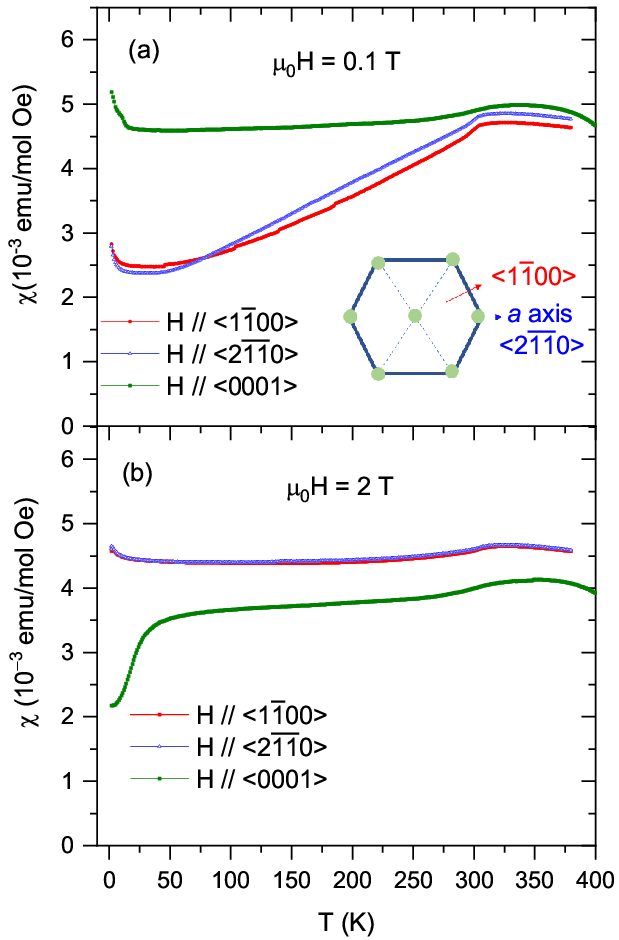}
    \caption{DC magnetic susceptibility $\chi(T)$ measured under a zero-field-cooled (ZFC) protocol with an applied magnetic field of (a) 0.1 T and (b) 2 T, for fields oriented parallel to the crystallographic $\braket{2\overline{1}\overline{1}0}$, $\braket{1\overline{1}00}$, and $\braket{0001}$ axes. Inset of panel (a) illustrates the orientation of the $\braket{2\overline{1}\overline{1}0}$ and $\braket{1\overline{1}00}$-axes within the hexagonal basal plane; the $\braket{0001}$ axis is oriented perpendicular to this plane.}
    \label{fig:MT}
\end{figure} 
\begin{figure}[htbp!]  
    \centering
    \includegraphics[clip,width=0.95\columnwidth]{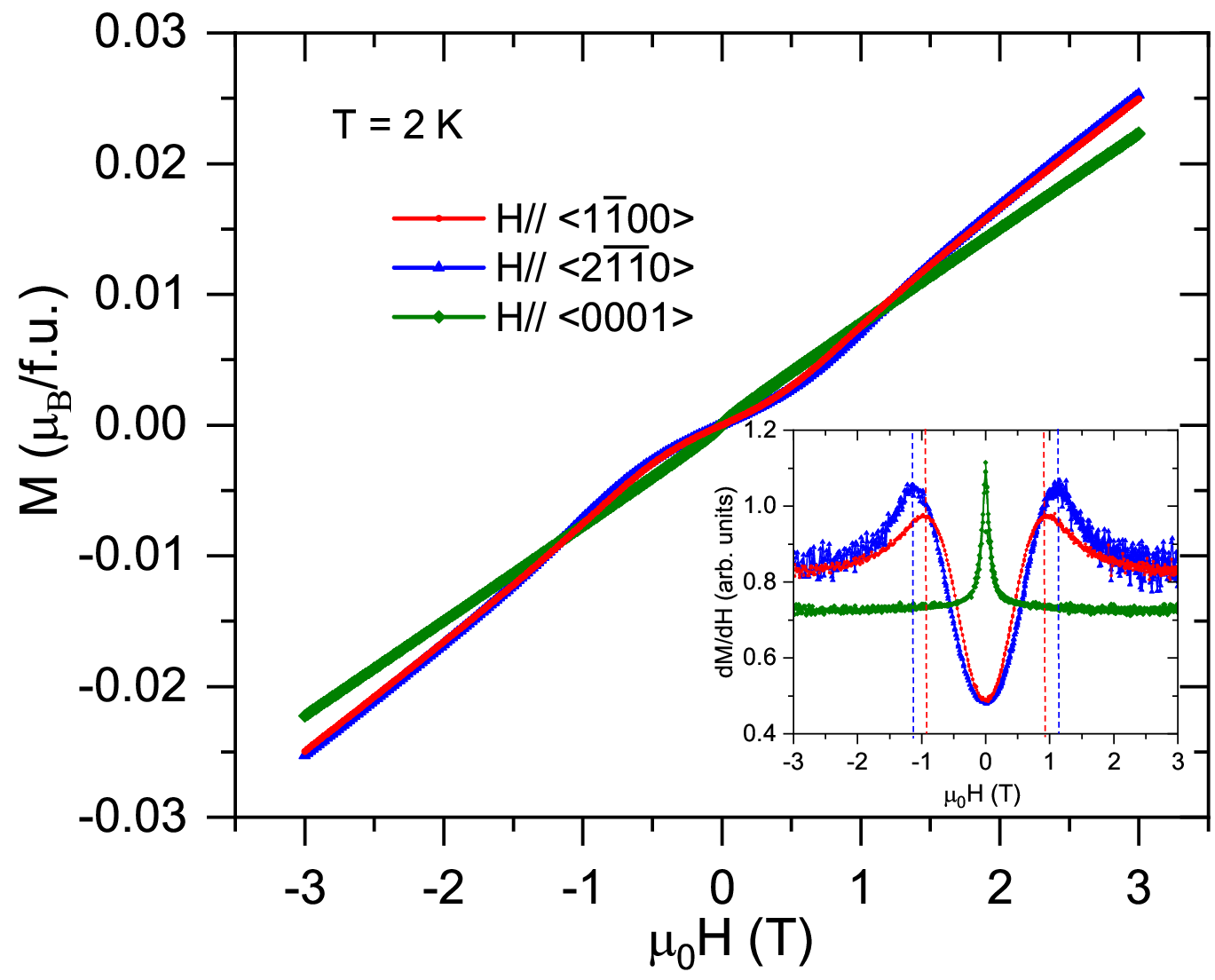}
    \caption{Magnetization $M(H)$ measured at 2\,K with the magnetic field applied parallel to the crystallographic $a$-, $m$-, and $c$- axes. The inset shows the field derivative $dM/dH$ as a function of $H$. Vertical dashed lines mark the critical fields for both positive and negative fields applied along the $\braket{2\overline{1}\overline{1}0}$- and $\braket{1\overline{1}00}$-axes. A sharp peak in the $dM/dH$ curve for $H\parallel c$ is attributed to weak ferromagnetism along the $c$-axis.}
    \label{fig:MH}
\end{figure}
The temperature dependence of the DC magnetic susceptibility $\chi(T)$ data, measured using a zero-field-cooling (ZFC) protocol in an applied magnetic field of 0.1\,T and 2\,T, are presented in Fig.\,\ref{fig:MT}(a,b) with the magnetic field applied along different crystallographic directions as illustrated in the inset of Fig.\,\ref{fig:MT}(a).  For in-plane field orientations, $\chi(T)$ exhibits a clear decrease at $T_\mathrm{N} = 307\,K$, signaling the onset of antiferromagnetic order. This transition temperature is consistent with values reported in the literature \cite{Kluc2024,Krem2024,Krieg2017}.
In contrast, for a magnetic field applied parallel to the $c$-axis, $\chi(T)$ shows a broad maximum centered around $T\,\approx\,338\,$K.
The susceptibility data are indicative of the easy-plane anisotropy below $T_\mathrm{N}$, whereas the in-plane susceptibility is almost isotropic, which is consistent consistent with previous reports \cite{Szus2006,Krieg2017}. The subtle differences observed in Fig.\,\ref{fig:MT}(a) can be attributed to unequal contributions from different magnetic domains \cite{Takegami2025}. This weak in-plane anisotropy is further corroborated by magnetization $M(H)$ measurements performed with the magnetic field aligned along the $\braket{2\overline{1}\overline{1}0}$ ($a$-axis), $\braket{1\overline{1}00}$ (axis at an angle $30^\circ$ to $a$-axis), and $\braket{0001}$ ($c$-axis), as shown in Fig.\,\ref{fig:MH}. The $M(H)$ curves for the $a$- and $\braket{1\overline{1}00}$-axes are nearly identical and exhibit a change in slope, which is generally associated with spin-flop transition \cite{Krieg2017,Orlova2025}. However, this anomaly is relatively weak compared to the sharp jumps typically observed in materials with strong uniaxial anisotropy \cite{Jakobs2025,Felcher1996}. The weak anomaly is actually originating from the metastable domains and is discussed in subsection B in detail. 

\begin{figure}[htbp!]  
    \centering
    \includegraphics[clip,width=0.95\columnwidth]{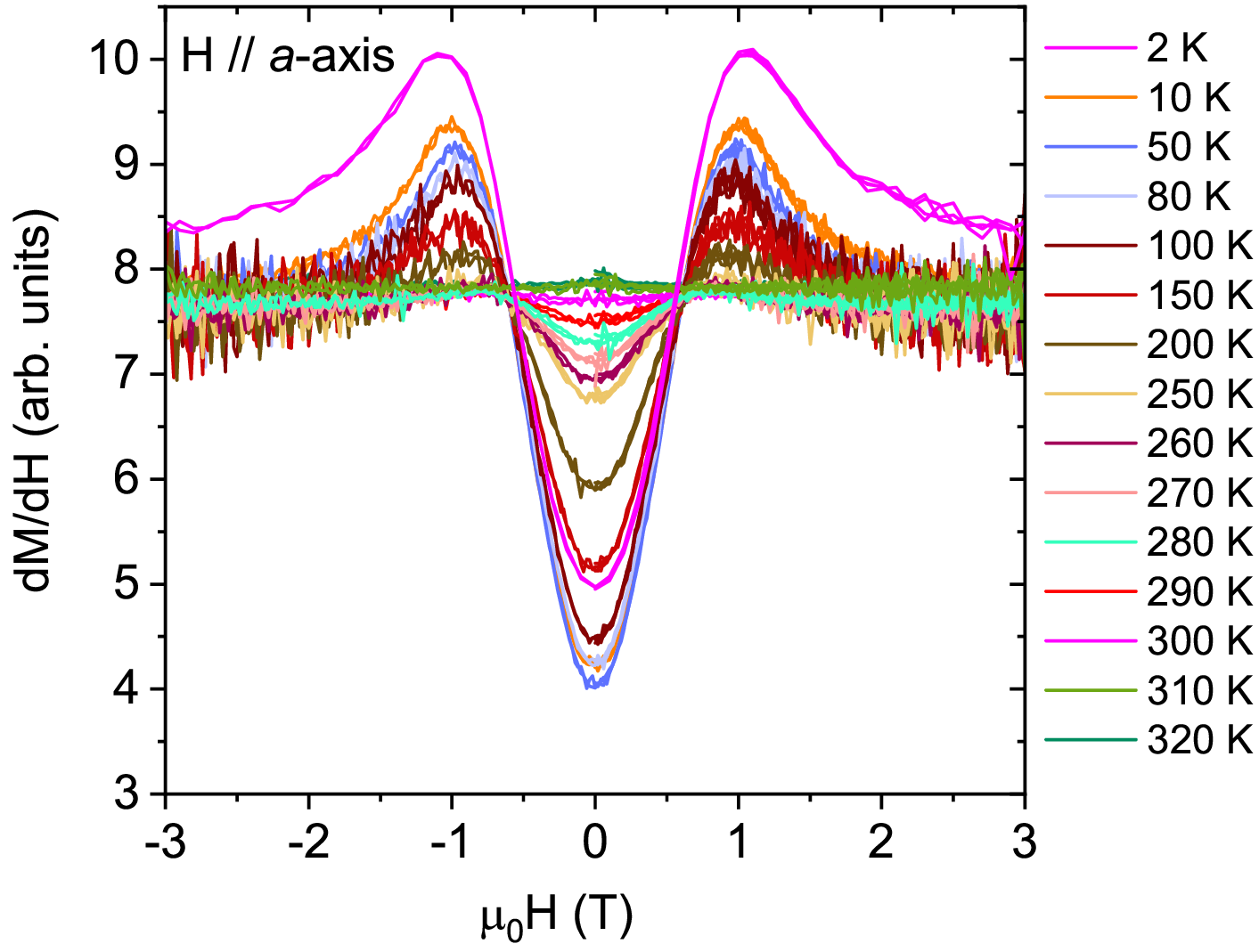}
    \caption{Magnetic field derivative $dM/dH$ of the magnetization $M(H)$ curves measured at various temperatures.}
    \label{fig:derivative}
\end{figure}

To quantify the critical field of this domain process $H_\mathrm{d}$, we analyzed the field derivative $dM/dH$, plotted in the inset of Fig.\,\ref{fig:MH}. The $H_\mathrm{d}$ was estimated from the peak-like features in the $dM/dH$ curves, marked by blue and red dashed lines for fields applied along the $a$- and $\braket{1\overline{1}00}$-axes, respectively. For the $a$ axis, the domain process occurs at $\mu_0 H_\mathrm{d} \approx$\,1.2\,T, whereas along the $\braket{1\overline{1}00}$ axis it occurs at a slightly lower field of $\mu_0 H_\mathrm{d} \approx$\,0.94\,T. No ferromagnetic component is observed in the 
$M(H)$ curves for field applied along $a$- and $\braket{1\overline{1}00}$ axes. However, for magnetic field applied parallel to the $c$ axis, a small ferromagnetic component is observed. This behavior is reminiscent of the weak ferromagnetism reported in previous studies \cite{Wasscher,Mazin2024,Kluc2024}. The sharp peak in the $dM/dH$ curve, shown in the inset of Fig.\,\ref{fig:MH}, corresponds to this weak ferromagnetic contribution. In the following, we examine the temperature dependence of both $H_\mathrm{d}$ and the weak ferromagnetism in more detail. Temperature-dependent field derivative $dM/dH$ of the magnetization $M(H)$ are presented in Fig.\,\ref{fig:derivative}. Each curve corresponds to a distinct measurement temperature. The $H_\mathrm{d}$ decreases systematically with increasing temperature and vanishes at the N\'eel temperature $T_\mathrm{N}$. This behavior highlights that the $H_\mathrm{d}$ is related to the antiferromagnetic order: its systematic reduction reflects the weakening of exchange anisotropy with increasing temperature, while its disappearance at $T_\mathrm{N}$ marks the loss of long-range antiferromagnetic order.

\begin{figure}[htbp!]  
    \centering
    \includegraphics[clip,width=0.95\columnwidth]{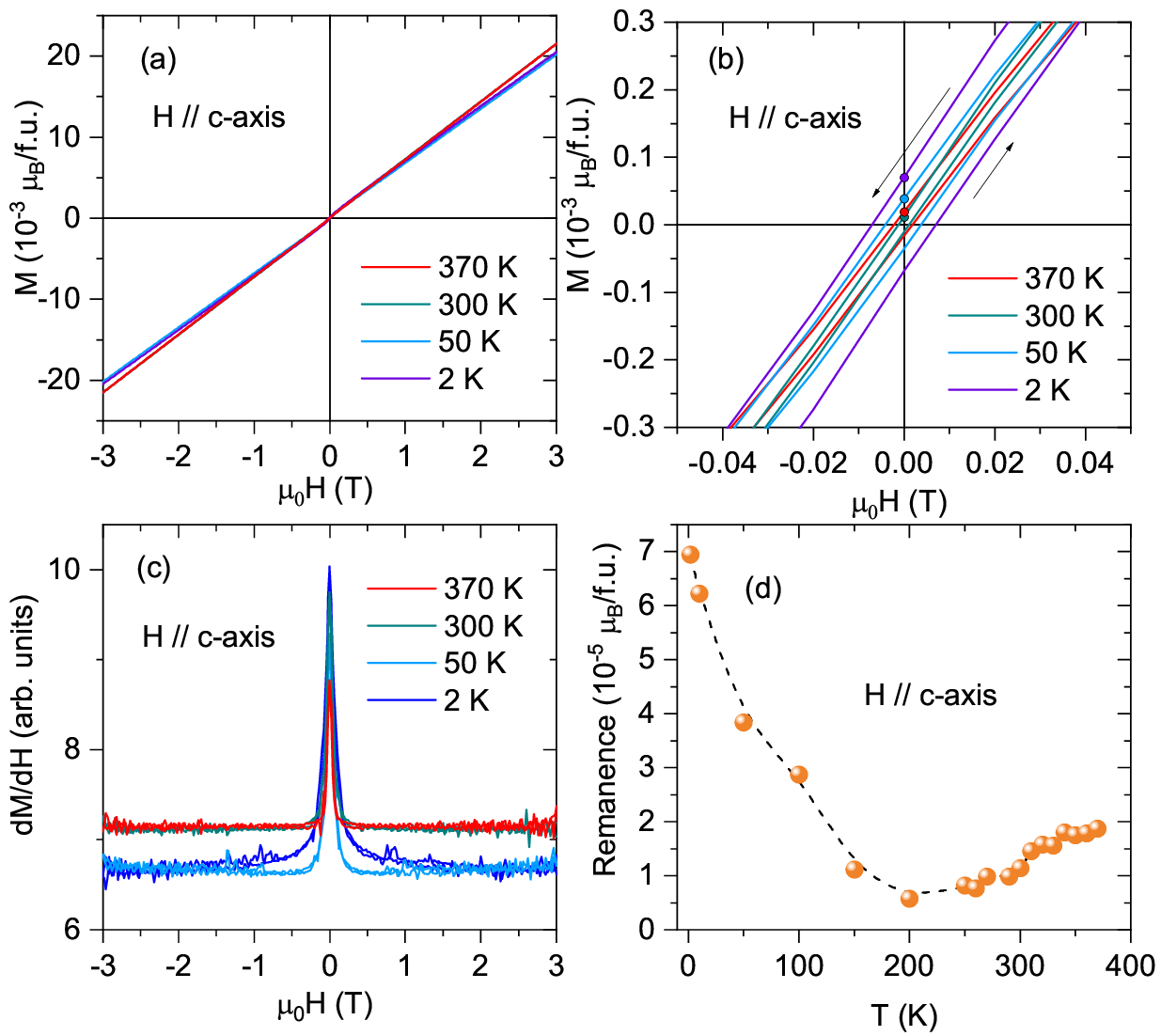}
    \caption{(a) Magnetic-field dependence of the magnetization $M(H)$ for selected temperatures above and below $T_\mathrm{N}$ with the crystallographic $c$-axis parallel to the applied field ($c \parallel H$). (b) Low-field zoom of (a) highlighting the hysteresis loop and finite remanence due to weak ferromagnetism. (c) Field derivative $dM/dH$ for the curves in (a). (d) Temperature dependence of the remanent magnetization.}
    \label{fig:c-axis}
\end{figure}

  In contrast, the $M(H)$ measurements performed in the temperature range 2--370\,K with the magnetic field applied $H\parallel\,c$ axis display a different trend as can be seen in Fig.\,\ref{fig:c-axis}. A deviation from linear behavior is observed in the $M(H)$ curves at low magnetic fields, as evident from Fig.\,\ref{fig:c-axis}(a) and (b). A finite hysteresis loop with remanent magnetization at low fields (Fig. \,\ref{fig:c-axis}(b)) confirms the weak ferromagnetism. The remanent magnetization is found to be $\approx 7 \times 10^{-5} \mu_\mathrm{B}$/f.u. at 2\, K, comparable to values reported previously \cite{Kluc2024,Mazin2024}. In contrast to Ref.\,\onlinecite{Kluc2024}, where the onset of weak ferromagnetism is shown to coincide with $T_\mathrm{N}$, our measurements reveal that weak ferromagnetism persists above $T_\mathrm{N}$, as evidenced by the field-derivative data presented for $T\,$=\,370\,K in Fig.\,\ref{fig:c-axis}(c). The temperature dependence of remanence field does not behave like an order parameter as shown in Fig. \ref{fig:c-axis}(d). The magnitude of remanence initially decreases with temperature below 370\,K followed by an increase below 200\,K, see Fig.~\ref{fig:c-axis}(d). The fact that neither the onset temperature nor the temperature at which the remanence begins to increase coincides with $T_\mathrm{N}$ indicates that the weak ferromagnetism is driven by defects. Intriguingly, the minimum in the remanent magnetization at 200-250 K coincides with the temperature range where a change in the magnetic structure was reported in the recent muon scattering experiment \cite{Hicken2025}. Similarly, in $\alpha$-MnTe thin films \cite{Amin2024}, the spontaneous anomalous Hall effect emerges at low temperatures and vanishes at higher temperatures, suggesting that it is linked to the strengthening of the ferromagnetic component below approximately 200\,K.
	
\begin{figure}[htbp!]  
    \centering
    \includegraphics[clip,width=0.95\columnwidth]{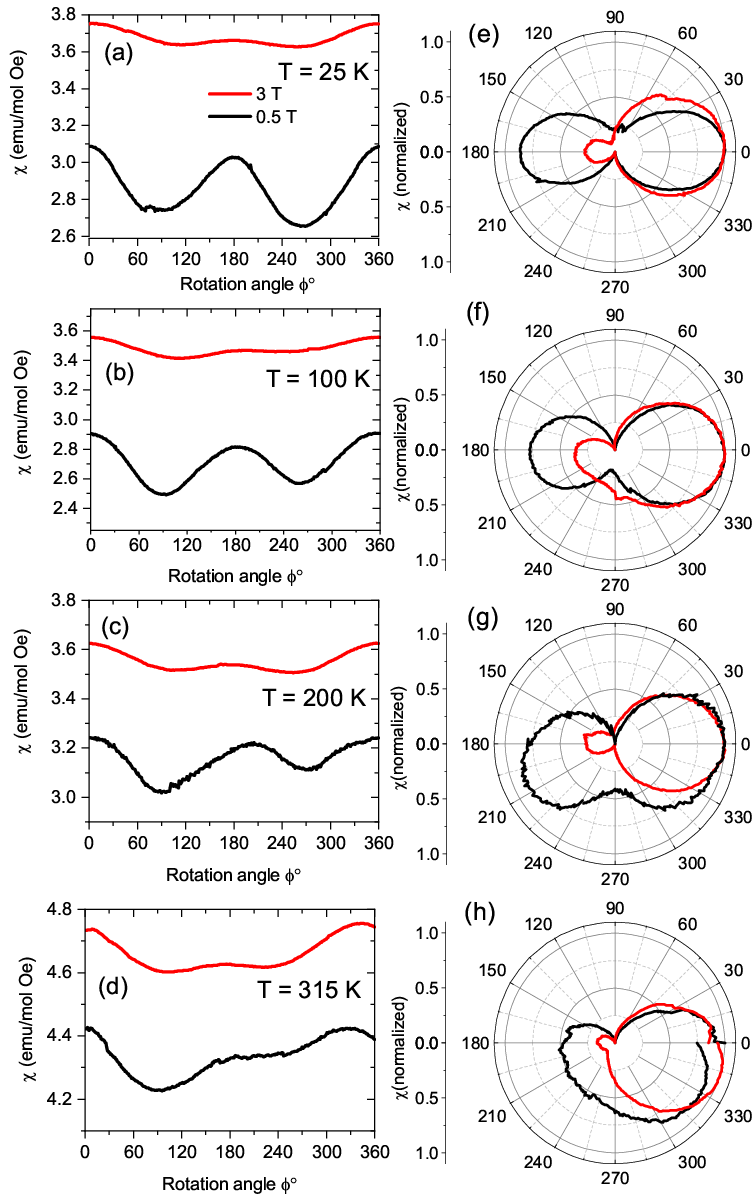}
    \caption{(a-d) Magnetic field angle $\phi$ dependence of dc-magnetic susceptibility $\chi(\phi)$ measured at different temperatures and magnetic fields of 0.5\,T and  3\,T. (e-h) corresponding normalized $\chi(\phi)$ presented in polar plots.}
		\label{Rotation}
\end{figure}

	In Fig. \ref{Rotation}(a–d), the in-plane angular dependence of the dc magnetic susceptibility, $\chi(\phi)$, is shown for different temperatures measured above and below $\mu_0 H_\mathrm{sf}$. The corresponding normalized $\chi(\phi)$ data are displayed as polar plots in Fig. \ref{Rotation}(e–h). Strikingly, for temperatures $T < T_\mathrm{N}$, the magnetic susceptibility $\chi(\phi)$ exhibits a $\pi$-periodic easy-axis anisotropy in a magnetic field of 0.5 T. This behavior is reminiscent of that observed in $\alpha$-MnTe crystals grown by the gradient-freeze method \cite{Orlova2025}. However, $90^\circ$ rotation of the easy axis as reported by Orlova $et~al.$ \cite{Orlova2025} below 81 K, was not observed in our crystals, consistent with the fact that we neither observe the onset of weak ferromagnetism below this temperature. The same $\pi$-symmetry is preserved up to $T_\mathrm{N}$ for low-field measurements as shown in Fig.~\ref{Rotation}. By contrast, the $\pi$-symmetry observed in $\chi(\phi)$ is lost when the magnetic field is increased to 3\,T, i.e., above the characteristic field $\mu_0 H_\mathrm{d}$. This behavior indicates that the magnetization vector can not be easily rotated in the $ab$-plane, and a complex multi-domain texture with contribution from higher-order anisotropy terms \cite{Bogdanov2007} are present in $\alpha$-MnTe. For $T > T_\mathrm{N}$, as expected, the two-fold symmetry is lost. However, the shape of $\chi(\phi)$ is retained for the 3\,T measurement, which indicates that the strong magnetic correlations persist above  $T_\mathrm{N}$.  
\begin{figure}[htbp!]  
    \centering
    \includegraphics[clip,width=0.95\columnwidth]{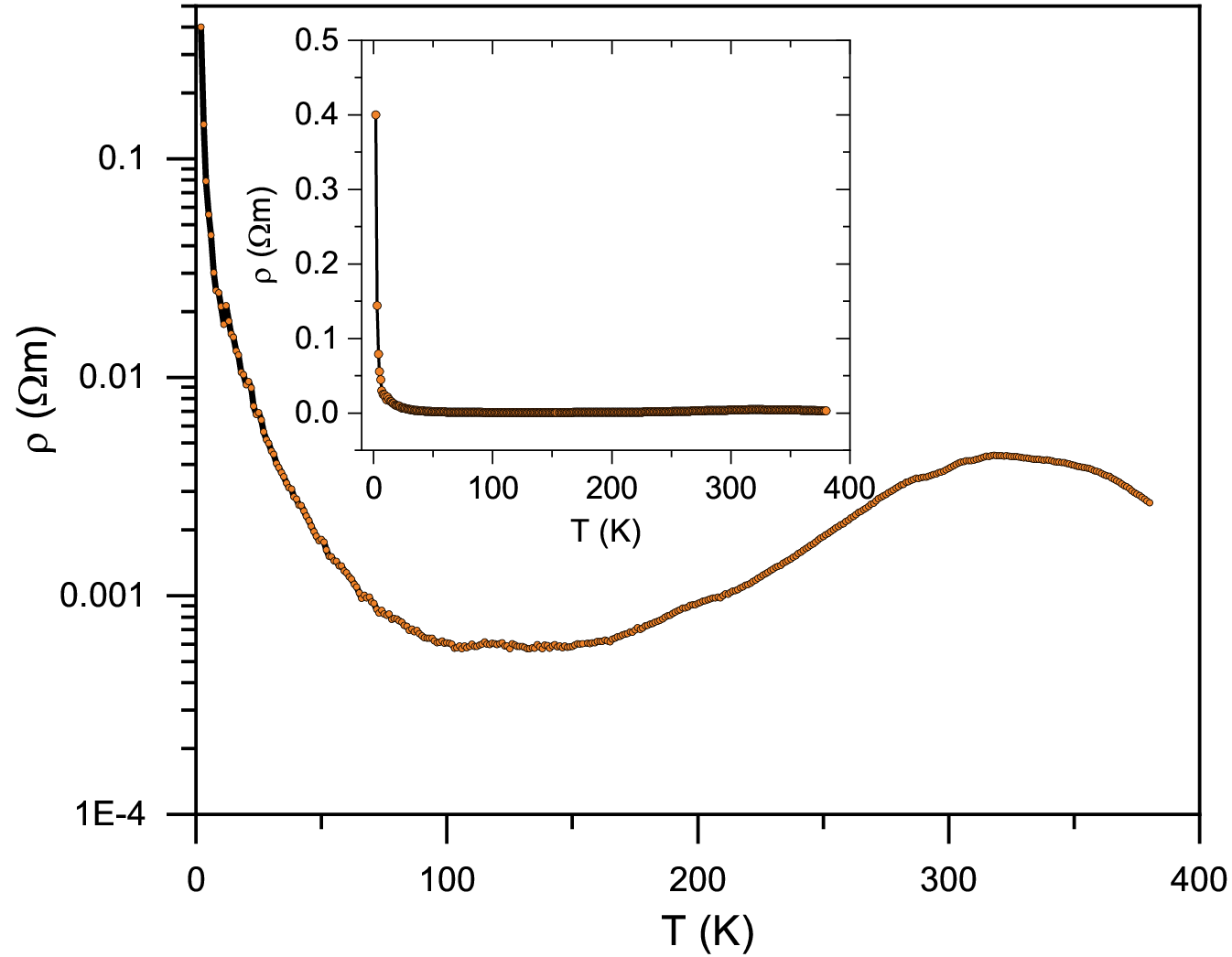}
    \caption{Resistivity $\rho(T)$ of a $\alpha$-MnTe single crystal plotted on the logarithmic scale. Same data is shown on the linear scale in the inset.}
		\label{RT}
\end{figure}
\begin{figure}[htbp!]  
    \centering
    \includegraphics[clip,width=0.95\columnwidth]{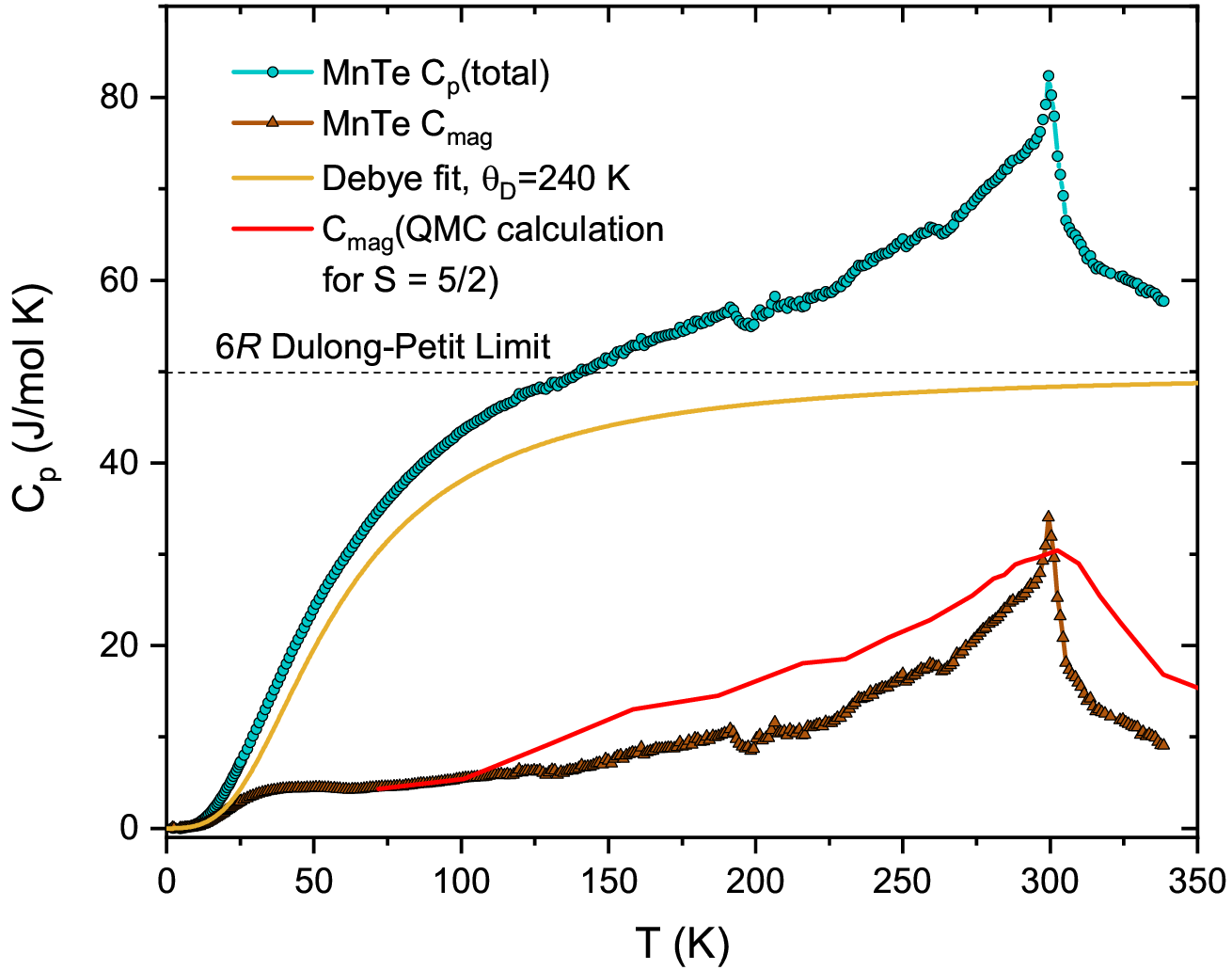}
    \caption{Temperature dependence of specific heat $C_p(T)$ of $\alpha$-MnTe single crystals. The magnetic specific heat for a $S\,=5/2$ system was calculated using a quantum Monte Carlo (QMC) method. The total experimental $C_p(T)$ can be described by a lattice part fitted using a Debye model and the magnetic specific heat calculated using QMC method for a $S\,=5/2$ system. The Debye model gives a Debye temperature $\theta_D\,=240\,K.$ }
		\label{HC}
\end{figure}

Figure \ref{RT} shows the temperature dependence of the resistivity $\rho(T)$ of $\alpha$-MnTe single crystals measured between 370 and 2 K on a logarithmic scale. The inset presents the same data on a linear scale. Below the Néel temperature $T_\mathrm{N}$, the resistivity decreases due to the suppression of spin–disorder scattering as the Mn spins order antiferromagnetically. At temperatures below ~100 K, however, the resistivity increases with decreasing temperature as carriers freeze out — a characteristic behavior of doped semiconductors. Similar $\rho(T)$ was reported for single crystals grown by self-flux technique \cite{Dzian2025,Wu2025}. In $\alpha$-MnTe, optical studies \cite{Allen1977,Roca2000} have detected a direct band gap of 1.3 eV, which would yield a resistivity in the range of 10$^{3}\Omega$m. The overall resistivity found for our $\alpha$-MnTe crystals lies within the typical range for doped semiconductors and varies by about two orders of magnitude between room temperature and 2 K.

In order to characterize the magnetic phase transition of $\alpha$-MnTe, we performed specific heat measurements $C_p(T)$ in the temperature range 2-338\,K. The peak corresponding to the antiferromagnetic transition is been at $T_\mathrm{N}\,\approx$\,300\,K, which is consistent with a previous report \cite{Gron1972}. No signature of any additional phase transition has been found in our samples below 300 K down to 2 K, thus ruling out any additional ferromagnetic phase transitions at lower temperatures. Since both phonon and and magnon excitations have a $T^3$ temperature dependence at low temperatures,  Debye (Eq.\ref{Debye}) fit overestimates phonon contribution and underestimates magnetic entropy. Since the neutron scattering experiments \cite{Kuni1964} on $\alpha$-MnTe detected magnetic moment corresponding to the $S = 5/2$ state, we used quantum Monte Carlo (QMC) simulations performed within the ALPS package \cite{Albu2007} to calculate the temperature dependence of the magnetic part of the specific heat for a $S = 5/2$ system, see Fig. \ref{HC}. Simulations were performed for the finite lattice with periodic boundary conditions and up to 1024 sites using exchange parameters from Ref. \onlinecite{Liu2024a}. The remaining part of the specific heat is described by the lattice part of the specific heat can be well-described by Eq.\ref{Debye}.         

\begin{equation} 
C_{\rm Ph}(\theta_{\rm D}, T) = n 9 R \left( \frac{T}{\theta_{\rm D}} \right)^3 
\int_0^{\theta_{\rm D}/T} \frac{x^4 e^x}{(e^x - 1)^2} \,dx
\label{Debye}
\end{equation}

where the number of atoms is n=2 for MnTe, R is the gas constant and $\theta_{\rm D}$ is the Debye temperature. By combining Eq.\ref{Debye} with the QMC data for the magnetic specific heat, we obtain $\theta_{\rm D}$ = 240 K for $\alpha$-MnTe.  

\subsection{Micromagnetic model}

As already mentioned in section A, there are anomalies in the magnetization $M(H)$ at low fields which have been associated with spin-flop transitions\cite{Krieg2017,Orlova2025}. These magnetization processes, however, do not agree with the expected 
behavior of a conventional spin-flop in an easy-axis antiferromagnet \cite{neel1952,Bogdanov2007}.
In principle, the magnetization process should reflect the hexagonal symmetry where any vanishingly small in-plane field
overturns the antiferromagnetic configuration such that magnetic moments are perpendicular or almost perpendicular
within 30$^{\circ}$ to the field. Magnetization processes in hexagonal antiferromagnets then should always involve domain processes related to moving domains walls, as described in Ref.\,\onlinecite{Bogdanov1998}. 
%
%
Moreover, unidirectional angular dependencies have been reported \cite{Orlova2025},
and similar results are found here for our crystals under rotating field.
These results are not conforming with 
the drastic angular dependence of real uniaxial magnetic system where 
the spin-flop region covers only a small
region in the general magnetic phase diagram with varying field directions \cite{Bogdanov2007},
owing to the weak effects of anisotropy compared to the strong exchange interactions.
The unidirectional angular dependencies observed here are unexpected for any magnetic system and indicate 
that underlying processes must involve metastable states. 

Here, we present the general phenomenological model for hexagonal antiferromagnets
with easy-plane anisotropy and the allowed DM terms that can 
generate weak ferromagnetic polarization due to spin-canting for the specific magnetic 
order in $\alpha$-MnTe. The model corroborates that magnetic anomalies indicating complex 
magnetization processes in these crystals rely on domain processes and do not involve 
jump-like first-order re-orientation of the staggered vector which would be common in 
easy-axis antiferromagnets.

The phenomenological free energy density for the hexagonal two-sublattice antiferromagnet $\alpha$-MnTe
enables a qualitative analysis of the magnetization processes in single crystals.
Its standard form in terms of the magnetizations $\mathbf{M}_i$ of the $i$-th sublattice includes isotropic
exchange, Zeeman energy for the internal magnetic field $\mathbf{H}$ and anisotropic terms $w_a$
\cite{Bogdanov1998}. Additionally, we consider terms from higher-order DM interactions \cite{Mazin2024}
\begin{equation}\label{w}
	w=\lambda\,\mathbf{M}_1 \cdot \mathbf{M}_2 - \mathbf{H} \cdot (\mathbf{M}_1 + \mathbf{M}_2) + w_a + w_{\Delta},
\end{equation}
where $\lambda$ is the effective exchange constant.

Application of model for temperatures much lower than the N{\'e}el temperature 
and in low magnetic fields allows to simplify the model by standard arguments.
The magnitude of the sublattice magnetization, $|\mathbf{M}_i|=M_0$
is fixed to the spontaneous magnetization $M_0$ that is a constant at a given temperature. 
Both anisotropy energy and DM terms are small, $w_a,w_{\Delta} \ll \lambda M_0^2$. 
Therefore, these terms can be written as invariants 
in powers of vector components from 
the staggered vector $\mathbf{l}$, while keeping only linear 
terms from the net ferromagnetic moment $\mathbf{m}$,
as in the Zeeman energy.
These two vectors are defined as 
\begin{equation}\label{landm}
	\mathbf{M}_{1,2}=M_0 (\mathbf{m} \pm \mathbf{l})
\end{equation}
and have the property $\mathbf{m} \cdot \mathbf{l} =0$, $\mathbf{m}^2 + \mathbf{l}^2=1$
because of the micromagnetic approximation $|\mathbf{M}_1|=|\mathbf{M}_2|=M_0$.
Then, the anisotropy energy is given by
\begin{equation}\label{wa}
	w_a (\mathbf{l})=K_1\, l_z^2 + K_2\, ( l_x^2+l_y^2)^2+\frac{K_3}{2}\,[(l_x+il_y)^6+(l_x-il_y)^6]
\end{equation}
with $K_i$ anisotropy coefficients. The first two terms describe the uniaxial magnetic energy of the hexagonal crystal. 
If $K_1>0$, then the antiferromagnet is an easy-plane system with the $xy$-plane as
the basal plane perpendicular to the $c$-axis along $z$. 
This is the case of $\alpha$-MnTe. The sixth-order term describes the 
anisotropy for the staggered vector in the basal plane. 
Here, six easy directions exist. 
Rotated by 30$^{\circ}$ around $c$, there is another set of six directions
for the intermediate and unstable orientations of the staggered vector.
Depending on the sign of $K_3$ this system of easy and intermediate axes in the the basal 
plane has either easy axis along $x$ or rotated by 30$^{\circ}$ from it, i.e. easy axes point either
along (1$\bar{1}$00) or along (2$\bar{1}$00) in the hexagonal crystal system.

The magnetic symmetry of the staggered vector $\mathbf{l}$ in $\alpha$-MnTe (space group P$6_3/mmc$, No.194,
with Mn on Wyckoff position 2$a$) is such that the $(l_x,l_y)$ transform 
as magnetic irreducible co-respresentation $m\Gamma^{5+}$, and $l_z$ as $m\Gamma^{4+}$,
while $\mathbf{m}$ transforms as the axial vector components, 
i.e.  $(m_y,-m_x)$  tranforms as $m\Gamma^{6+}$ and $m_z$ as $m\Gamma^{2+}$.
Using these transformation properties, 
the leading invariants linear in vector components of $\mathbf{m}$ 
can be derived \cite{ISOTROPY},
\begin{eqnarray}
\begin{aligned}
	w_{\Delta}(\mathbf{m},\mathbf{l}) &=& \Delta_{ab}\,[l_x l_y l_z m_x + (l_x^2 - l_y^2)\,l_z)\,m_y] \\
	      &&~~+ \Delta_c\, (3\,l_y^2\,l_x-l_y^3)\,m_z  \,.
\end{aligned}
\label{wDelta}
\end{eqnarray}
with $\Delta_{ab}$ and $\Delta_{c}$ being material-dependent constants for higher-order DM couplings. In principle, 
a quantification of the $w_{\Delta}$ terms could be derived from the generalized bilinear Heisenberg model
proposed in Ref.~\onlinecite{Mazin2024}. However, the values for the microscopic coupling constants
are unknown as of now., and the general phenomenological model features two independent invariants 
corresponding to parameters $\Delta_{ab}$ and $\Delta_{c}$, 
as also pointed out by Mostovoy \cite{Mostovoy2025}.

With these expressions, the free energy density can be written in 
components of $\mathbf{l}$ and $\mathbf{m}$ 
\begin{equation}\label{wml}
	w(\mathbf{l},\mathbf{m})=2\,\lambda M_0^2 \mathbf{m}^2 - 2 M_0 \mathbf{H} \cdot \mathbf{m} + w_a(\mathbf{l})+w_{\Delta}(\mathbf{l},\mathbf{m})\,.
\end{equation}
An independent minimization of the potential function (\ref{wml}) in $\mathbf{m}$, 
obeying $\mathbf{m} \cdot \mathbf{l}=0$, gives the net magnetization 
in terms of the staggered vector $\mathbf{l}$ 
\begin{equation}\label{mofl}
	\mathbf{m}=\frac{1}{4\lambda M_0} [\mathbf{H}-(\mathbf{H}\cdot\mathbf{l})/|\mathbf{l}|^2]+\mathbf{f}
\end{equation}
where the  vector of weak ferromagnetic moment (wfm) $\mathbf{f}$ is 
\begin{equation}\label{lwfm}
	\mathbf{f}=\mathbf{f}_0+g(\mathbf{l})\,\mathbf{l}
\end{equation}
with the first contribution in components of vector  $\mathbf{f}_0$ with 
Cartesian components,
\begin{eqnarray}
\begin{aligned}
	f_{0x} &=& - \frac{1}{8\lambda M_0^2}\,[\Delta_{ab}\,l_x l_y l_z] \\ 
	f_{0y} &=&- \frac{1}{8\lambda M_0^2}\,[\Delta_{ab}\,(l_x^2 - l_y^2) l_z] \\ 
	f_{0z} &=&- \frac{1}{8\lambda M_0^2}\,[\Delta_{c}\,(3l_y^2 l_x - l_y^3)] \,,
\end{aligned}
\label{fwfm}
\end{eqnarray}
and the second contribution defined by 
the function  $g(\mathbf{l})$ in components of the staggered vector,
\begin{equation}\label{g}
g(\mathbf{l})=\frac{1}{8\lambda M_0^2}\,[\Delta_{ab}\,(2l_x^2 l_y l_z - l_y^3 l_z) + \Delta_{c}\,(3l_y^2 l_x - l_y^3) l_z]/|\mathbf{l}|^2
\,.
\end{equation}
This term corrects the spontaneous term  $\mathbf{f}_0$ in order to respect the condition $\mathbf{m}\cdot\mathbf{l}=0$.
According to Eqs.~(\ref{mofl}) and (\ref{fwfm}), weak ferromagnetic moments $\mathbf{f}$ 
from a canting of the sublattices appear for general orientation of the staggered vector 
as a spontaneous effect in zero field $\mathbf{H}=0$.
Assuming an orientation of the antiferromagnetic vector strictly in the basal plane,
$l_z \equiv 0$, we have $g(\mathbf{l})=0$ and also $f_{x,y} \equiv 0$. Hence, a net magnetization 
is found only along the $c$-axis,  $f_z = - (1)/(8\lambda M_0)\,\Delta_{c}\,(3l_y^2 l_x - l_y^3)$.
In this case with strong easy-plane anisotropy, applying magnetic fields in directions of the basal plane, 
and generally in weak magnetic fields, this spontaneous weak ferromagnetism will dominate the net polarization of the system.
Reinserting the expression for the net magnetization (\ref{mofl}) into the free energy (\ref{wml}) 
gives the thermodynamic potential in terms of the staggered vector 
\begin{eqnarray}
\begin{aligned}
	\tilde{w}(\mathbf{l})  = & \frac{1}{8\lambda}\left[\mathbf{H}\cdot\mathbf{H}-\frac{ (\mathbf{H}\cdot\mathbf{l})^2}{|\mathbf{l}|^2}\right] & \,\\
	 + & \frac{1}{2\lambda M_0}\left[\mathbf{H}-\frac{ (\mathbf{H}\cdot\mathbf{l})\,\mathbf{l}}{|\mathbf{l}|^2}\right] 
\,\cdot\,\mathbf{f} &\,\\
	 +  & (\mathbf{f}\,\cdot\,\mathbf{f})+w_a(\mathbf{l}) &\,.
\end{aligned}
\label{wtilde}
\end{eqnarray}
The magnetic thermodynamic potential is now a function of $\mathbf{l}$ alone. The terms in the two last lines for $\tilde{w}$ in Eq.~(\ref{wtilde}) define a generalized anisotropy including the effects of the DM-terms.

Representing the staggered vector in spherical coordinates,\,
\(\mathbf{l}=l\,(\cos(\varphi),\sin(\vartheta),\sin(\varphi)\sin(\vartheta),\cos(\vartheta))\), and using the expression for the net magnetization, Eq.~(\ref{mofl})-(\ref{g}), the condition on the magnitude of the sublattice magnetizations $|\mathbf{M}_{1,2}|\equiv M_0$ from the  micromagnetic approximation Eq.~(\ref{landm}) gives a polynomial equation for the modulus $l$ of the staggered vector. 
%
The polynomial equation for the modulus $l$ of the staggered vector 
can be written
\(\sum_{i=0}^{6}\,{\alpha}_{i}\,l^{i}  = 0\)
with the coefficents given by functions of the angular variables, 
$\alpha_1=\alpha_4=\alpha_5=0$
and
\begin{eqnarray}
\alpha_0&=&
4  M_0 ^2 \left(\left(H_x^2+H_y^2-\lambda^2  M_0 ^2\right) \cos[\vartheta]^2 \right. \\
	&\ & \hspace{24pt} +\frac{1}{2} \left(H_x^2+H_y^2+2 H_z^2-2 \lambda^2  M_0 ^2+ \right. \nonumber\\
	&\ & \hspace{40pt} \left(-H_x^2+H_y^2\right) \cos[2 \varphi] \nonumber\\
	&\ & \hspace{42pt} \left. -2 H_x H_y \sin[2 \varphi]\right) \sin[\vartheta]^2 \nonumber \\
	&\ & \hspace{42pt} -H_z (H_x \cos[\varphi]+H_y \sin[\varphi]) \nonumber\\
	&\ & \hspace{48pt} \left. \sin[2 \vartheta]\right) \,, \nonumber
\end{eqnarray}
%
%
\begin{eqnarray}
\alpha_{2}& =& 4 \lambda^2 M_0 ^4 
\end{eqnarray}
\begin{eqnarray}
	\alpha_3 & = & 	-4 \Delta_{ab}  M_0  \sin[\vartheta]^2 \left(H_y \cos[\varphi]^4 \cos[\vartheta] \sin[\vartheta]^2 \right. \\
	&\ & \hspace{24pt} +\cos[\varphi]^3 \sin[\varphi] \sin[\vartheta]^2 \nonumber \\
	&\ & \hspace{12pt} (-H_x \cos[\vartheta]+3 \Delta_{c} H_z \sin[\varphi] \sin[\vartheta]) \nonumber\\
	&\ & \hspace{12pt} -\sin[\varphi]^2 (H_y \cos[\vartheta]-H_z \sin[\varphi] \sin[\vartheta]) \nonumber\\
	&\ & \hspace{12pt} \left(\cos[\vartheta]^2-\Delta_{c} \sin[\varphi]^2 \sin[\vartheta]^2\right) \nonumber\\ 
	&\ & \hspace{12pt} +\cos[\varphi]^2 \left(H_y \cos[\vartheta]^3-2 H_z \cos[\vartheta]^2 \sin[\varphi] \sin[\vartheta] \right. \nonumber \\
	&\ & \hspace{12pt}  -(3 \Delta_{c} H_x+2 H_y) \cos[\vartheta] \sin[\varphi]^2 \sin[\vartheta]^2 \nonumber \\
	&\ & \hspace{12pt} \left. -\Delta_{c} H_z \sin[\varphi]^3 \sin[\vartheta]^3\right)+\cos[\varphi] \nonumber\\
	&\ & \hspace{12pt} \left(H_x \cos[\vartheta]^3 \sin[\varphi] \right. \nonumber \\
	&\ & \hspace{12pt} \Delta_{c} (H_x-3 H_y) \cos[\vartheta] \sin[\varphi]^3 \sin[\vartheta]^2+\sin[\varphi]^3 \sin[\vartheta] \nonumber \\
	&\ & \hspace{12pt} \left. \left. \left(3 \Delta_{c} H_z \sin[\varphi] \sin[\vartheta]^2+H_x \sin[2 \vartheta]\right)\right)\right) \,, \nonumber
\end{eqnarray}
and
\begin{eqnarray}
\alpha_6 & = & \frac{1}{8}
\Delta_{ab}^2 \sin[\vartheta]^4 \left((5+3 \cos[4 \varphi]) \cos[\vartheta]^4 \right. \\
	&\ &  \hspace{12pt} +8 \cos[\vartheta]^2 \left(\cos[\varphi]^6+6 \Delta_{c} \cos[\varphi] \sin[\varphi]^5 \right. \nonumber \\
	&\ &  \hspace{20pt} \left. -2 \Delta_{c} \sin[\varphi]^6\right) \sin[\vartheta]^2 \nonumber \\
	&\ &  \hspace{12pt} +8 \Delta_{c}^2 \sin[\varphi]^4 \nonumber \\
	&\ &  \hspace{12pt} (-3 \cos[\varphi]+\sin[\varphi])^2 \sin[\vartheta]^4+\sin[2 \varphi]^2 \nonumber \\
	&\ &  \hspace{12pt} \left. (\Delta_{c}-(2+\Delta_{c}) \cos[2 \varphi]-3 \Delta_{c} \sin[2 \varphi]) \sin[2 \vartheta]^2\right) \,. \nonumber
\end{eqnarray}
%
%
There is one real positive root of this equation $l=l(\varphi,\vartheta)$ with $0\leq l \leq 1$,  which describes the relevant solutions of the staggered vector. Reinserting this expression into the potential (\ref{wtilde}) yields it in the convenient form of an expression $w=w(\varphi,\vartheta)$, depending on angular variables alone. This expression can be evaluated numerically with arbitrary precision. By setting $\lambda=1$ and $M_0=1$ the potential can be represented in non-dimensional form 
%
as the sixth order 
polynomial in terms of the function $l=l(\varphi,\vartheta)$, 
\begin{math}\label{Phi}
\Phi(\varphi,\vartheta)=\sum_{i=0}^6\,c_{i}\,l^{i}\,
\end{math}
with coefficients as functions of the angular variables.
The coefficients are $c_1=c_5=0$ and
\begin{eqnarray}
c_0&=&-\left(h_x^2+h_y^2\right) \cos[\vartheta]^2 \\
   &\ & -\frac{1}{2} \left(h_x^2+h_y^2+2 h_z^2+\left(h_y^2-h_x^2\right) \cos[2 \varphi] \right. \nonumber \\
   &\ & \hspace{20pt} \left. -2 h_x h_y \sin[2 \varphi]\right) \sin[\vartheta]^2 \nonumber \\
   &\ & \hspace{4pt} +h_z (h_x \cos[\varphi]+h_y \sin[\varphi]) \sin[2 \vartheta] \,, \nonumber
\end{eqnarray}
\begin{equation}
	c_2=
\frac{1}{2} (-2+k_1+k_1 \cos[2 \vartheta])
\end{equation}
\begin{eqnarray}
c_3 &=&
\frac{1}{8} \Delta_{ab} \sin[\vartheta]^2 \left[2 \cos[2 \varphi] \right. \\
   &\ & \hspace{20pt} (h_y \cos[\vartheta] (3-\Delta_{c}+(1+\Delta_{c}) \cos[2 \vartheta]) \nonumber\\
   &\ & \hspace{12pt} +h_z (-3+\Delta_{c}-(3+\Delta_{c}) \cos[2 \vartheta]) \sin[\varphi] \sin[\vartheta])  \nonumber \\
	&\ & \hspace{12pt} +2 h_z \sin[\varphi] \sin[\vartheta] \left(-1-\Delta_{c}+(-1+\Delta_{c}) \cos[2 \vartheta] \right. \nonumber\\
	&\ & \left. \hspace{38pt} +6 \Delta_{c} \sin[2 \varphi] \sin[\vartheta]^2\right)  \nonumber \\
	&\ & \hspace{12pt} +\cos[\vartheta] \left(((3+\Delta_{c}) h_x-3 \Delta_{c} h_y \right. \nonumber \\
	&\ & \hspace{32pt}  +(h_x-\Delta_{c} h_x+3 \Delta_{c} h_y) \cos[2 \vartheta]) \sin[2 \varphi] \nonumber \\
	&\ & \hspace{32pt}  +(-3 \Delta_{c} h_x+h_y \nonumber \\
	&\ &  \hspace{32pt}  +3 \Delta_{c} h_y+(3 \Delta_{c} h_x+(3+\Delta_{c}) h_y) \cos[4 \varphi] \nonumber \\
	&\ &  \hspace{32pt} \left.\left. -((3+\Delta_{c}) h_x-3 \Delta_{c} h_y) \sin[4 \varphi]) \sin[\vartheta]^2\right)\right] \,, \nonumber
\end{eqnarray}
\begin{equation}
        c_4= k_2 \sin[\vartheta]^4 \,,
\end{equation}
and
\begin{eqnarray}
	c_6&=&
\frac{1}{256}
\left(-16 \Delta_{ab}^2 (3+\cos[4 \varphi]) \cos[\vartheta]^4 \sin[\vartheta]^4 \right. \\
	&\ & \hspace{32pt} -2 \cos[\vartheta]^2 \left(2 (5-6 \Delta_{c}) \Delta_{ab}^2 \right. \nonumber \\
	&\ & \hspace{32pt} +(7+26 \Delta_{c}) \Delta_{ab}^2 \cos[2 \varphi] \nonumber \\
	&\ & \hspace{32pt} +2 (3-10 \Delta_{c}) \Delta_{ab}^2 \cos[4 \varphi] \nonumber\\
	&\ & \hspace{32pt} +\left((9+6 \Delta_{c}) \Delta_{ab}^2-128 k_3\right) \cos[6 \varphi] \nonumber\\
	&\ & \hspace{32pt} -6 \Delta_{c} \Delta_{ab}^2 \sin[2 \varphi]-24 \Delta_{c} \Delta_{ab}^2 \sin[4 \varphi] \nonumber \\
	&\ & \hspace{32pt} \left. +18 \Delta_{c} \Delta_{ab}^2 \sin[6 \varphi]\right) \sin[\vartheta]^6 \nonumber\\
	&\ & \hspace{32pt} +64 \left(4 k_3 \cos[\varphi]^6-60 k_3 \cos[\varphi]^4 \sin[\varphi]^2 \right. \nonumber\\
	&\ & \hspace{32pt} +3 \left(-3 \Delta_{c}^2 \Delta_{ab}^2 +20 k_3\right) \cos[\varphi]^2 \sin[\varphi]^4 \nonumber\\
	&\ & \hspace{32pt} +6 \Delta_{c}^2 \Delta_{ab}^2 \cos[\varphi] \sin[\varphi]^5 \nonumber\\
	&\ & \hspace{32pt} \left. -\left(\Delta_{c}^2 \Delta_{ab}^2+4 k_3\right) \sin[\varphi]^6\right) \sin[\vartheta]^8 \nonumber\\
	&\ & \hspace{32pt} \left. +\Delta_{ab}^2 \sin[2 \varphi]^2 \sin[2 \vartheta]^4\right) \,. \nonumber
\end{eqnarray}
%
%
A complete analysis of this potential, depending on the internal magnetic field $\mathbf{h}=(h_x,h_y,h_z)$ and materials parameters for anisotropy $k_1$, $k_2$, $k_3$ and DM interactions $\Delta_{ab}$, $\Delta_c$ is a formidable task beyond scope of this work.  And, as crucial parameters for $\alpha$-MnTe are unknown as of now, here the discussion is restricted to qualitative pictures based on the clear hierarchy of the material-dependent magnetic parameters. The uniaxial in-plane anisotropy $k_1$ is strongest, but weak compared to exchange, therefore $0 < k_1 \ll 1$. As of now, effects of $k_2$ are unknown and, for the moment, we set $k_2=0$. The in-plane anisotropy is known to be very weak, $|k_3| \ll k_1$ in $\alpha$-MnTe \cite{Krieg2017,Povarov2025}. However, the DM interaction parameters could be relatively large, and we may assume $|k_3| < |\Delta_{ab}|,|\Delta_{c}| < k_1$. 
With these plausible assumptions, the behavior of the antiferromagnetism 
in $\alpha$-MnTe according to the general phenomenological model can be illustrated.

First, in zero field, there are the 
six orientations of the staggered vector in the plane. Thus, in zero internal magnetic field
a domain structure composed of three orientations of staggered vectors are possible. Each of these orientations allows the plus or minus orientation of the vector. Thus, regions with one orientation of the staggered vector are composed from two antiferromagnetic domains, separated by conventional 180$^{\circ}$ domain walls. Due to the presence of the DM interactions, there is a spontaneous weak ferromagnetic moment given by the vector $\mathbf{f}$, Eq.~(\ref{lwfm}). 
In this case, with the staggered vectors of the domains strictly in-plane $l_z=0$, i.e. $\vartheta=\pi/2$, the wfm has only components in $z$-direction.  This spontaneous moment along the crystallographic $c$ axis implies a slight canting of sublattice magnetization out of the basal plane. It depends on the in-plane orientation of the staggered vector as $f_{0z} \propto  -\cos(3\,\varphi)$. This means that two 180$^{\circ}$ domains with common orientation of the staggered vector have opposite up and down weak ferromagnetic polarization.

Under influence of the dipolar stray fields and magnetoelastic effects, domain structures composed of three orientations of staggered vectors are possible. Each of these orientations allows for plus or minus orientation of the staggered vector, separated by conventional 180$^{\circ}$ domain walls. The three different orientations of antiferromagnetic domain imply the existence of 60$^{\circ}$ and 120$^{\circ}$ domain walls. These walls are complicated owing to the magnetoelastic interactions, as magnetostriction distorts each of these domains into orthorhombic crystallographic twins. The co-existence of such domains as equilibrium states is possible, however, these macroscopic effects depend on shape and stress-state at the surface of finite samples \cite{Gomonay2002}. 
Within the micromagnetic model discussed here, co-existence of different antiferromagnetic domains in 
equilibrium can occur only in zero internal field and stress states, or for the specific case
with field in one easy axis direction and/or specific stress conditions such that different domains have
the same free energy. All other domain states can exist only in non-equilibrium, when metastable domains are kinetically arrested.

In arbitrarily small fields applied in the plane $(h_x,h_y,0)$ with general direction, only one of these orientations for the staggered vector forms the ground state. Thus, there are only two antiferromagnetic domains with $\pm$-direction. Only for orientation of the field along one of the easy axes, the two orientations of staggered fields along the other easy axes with totally four different antiferromagnetic domains can co-exist.
In the magnetic phase diagram for the hexagonal antiferromagnet with easy-plane anisotropy, there are no spin-flop transitions. This means, that there are no field-driven first-order transitions between equilibrium states for different orientations of the staggered vectors. This statement applies also in absence of DM interactions $\Delta_{ab}=\Delta_{c}=0$, as has been previously found \cite{Bogdanov1998}. 
Finally, it is noteworthy that the six magnetic antiferromagnetic states described by the potential $\Phi(\varphi,\vartheta)$ remain metastable in all in-plane applied fields with the above assumptions on the relative strength of anisotropies in $\alpha$-MnTe.

Thus, anomalous magnetization processes in a hexagonal antiferromagnet like $\alpha$-MnTe generically involve metastable domains and the field-driven re-distribution of domain states. These processes will always be observed, at least as transient states, following the argument by Bogdanov \cite{Bogdanov1998}, since a finite domain cannot simply disappear.
The interpretation of the anomalous magnetization curves for different field-orientations
in Fig.~\ref{fig:MH}, therefore, must not rely on the 
usual jump-like flopping of the staggered vectors within a homogenous magnetic state or domain. Instead, they must reflect complicated antiferromagnetic domain processes with redistribution of metastable domains, likely by motion of 60$^{\circ}$ and 120$^{\circ}$ domain walls.
Our observations, Fig.~\ref{Rotation} that the magnetization under rotating in-plane field does break the expected 180$^{\circ}$ periodicity and also displays curves which are not closed after 360$^{\circ}$ rotation conforms with such effect involving metastable states, as the magnetic state in these experiments changes in an irreversible manner by modification of the domain population.
Earlier experiments have displayed similar hysteretic effects, e.g. angular effects on magnetic response has been interpreted as lower crystallographic symmetry in Ref.~\cite{Orlova2025}. Clear hysteretic magnetization processes have been reported in Ref.~\cite{Beta2024}, cf. Fig.~S1, 
where $\alpha$-MnTe sample in a magnetoresistive device shows a clear relative shift of the response versus angle 
under right- and left-rotating field. These effects can only be explained by 
existence of metastable domains.

Returning to the micromagnetic model, the behavior in applied fields $h=(0,0,h_z)$ along the $c$-axis can 
be explained by the full polarization of the domains with common orientation. 
Our observation of a remanent state is explained by a disbalance among the 180$^{\circ}$ domains 
with common orientation of the staggered vector. 
For the ideal case of a full polarization of these domains, 
the expression for the wfm yields an estimate of $10^{-4}$ for the ratio of the DM interaction parameter $\Delta_{c}$ 
to exchange energy $\lambda\,M_0^2$, assuming a sublattice magnetization of $M_0\simeq 5\,\mu_B/\text{f.u.}$
using the obverved remanence of $7 \cdot 10^{-5}\,\mu_B/\text{f.u.}$ at low temperature. 
This estimate is a lower bound. Its magnitude is reasonable for DM interactions 
being in the range of $\mu$eV per bond, if the effective antiferromagnetic 
exchange is in the 10-100~meV range.
However, the estimate could deviate strongly from reality, 
even by orders of magnitude, as coercivity mechanisms 
for the motion of 180$^{\circ}$ walls are unkown in $\alpha$-MnTe.
Moreover, the remanence could derive from extrinsic, defect-related mechanisms, as
it persists in part at high temperatures above $T_{N}$ in these single crystals.

Moreover, the remanence may arise from extrinsic, defect-related mechanisms, as it persists even above $T_{N}$
in these single crystals. 
This remanence could be caused by Mn-interstitials or other point defects with large
spin-moment which behave as in a dilute magnetic semiconductor with a high effective Curie 
temperature but small net magnetization. We point out that the phenomenology of $\alpha$-MnTe in 
this case should be extended by considering a second ferromagnetic order-parameter  $\mathbf{F}$.
Coupling between the staggered vector of the antiferromagnetic order in $\alpha$-MnTe and this 
ferromagnetic order involves then terms of the form (\ref{wDelta}) with net 
magnetization $\mathbf{m}$ replaced by $\mathbf{F}$. 
The behavior of the antiferromagnetic order in $\alpha$-MnTe in each 
ferromagnetic domain then resembles 
that of the antiferromagnetic one in an internal 
magnetic field.

\section{Conclusions}

We synthesized $\alpha$-MnTe single crystals using iodine vapor transport and characterized them by x-ray diffraction, magnetization, resistivity, and specific-heat measurements. Synchrotron x-ray diffraction confirmed that the sample crystallizes in the P6$_3$/mmc space group. The physical characterization revealed an antiferromagnetic transition at $T_\mathrm{N} \approx 307$ K. A weak ferromagnetic signal was observed with the magnetic field applied parallel to the $c$-axis even above $T_\mathrm{N}$, and it becomes more pronounced below 200 K. The magnetic-field and angular dependence of the magnetization exhibited uniaxial and irreversible behaviors, which are unexpected for crystals with hexagonal symmetry. Using a phenomenological model, we show that the anomalies observed in magnetic fields originate from metastable domain configurations and irreversible magnetization processes.   


\begin{acknowledgments}
We gratefully acknowledge V. Hasse for assistance with the crystal growth and U. Nitzsche for support with the computational work. We thank ESRF for providing the beamtime for this experiment.

\color{black}

\end{acknowledgments}

\end{document}